\documentclass[12pt]{JHEP3}
\usepackage{amsmath, amssymb}
\usepackage{epsfig}
\usepackage{soul}

\newcommand{\beq}{\begin{equation}}
\newcommand{\eeq}{\end{equation}}
\newcommand{\baq}{\begin{eqnarray}}
\newcommand{\eaq}{\end{eqnarray}}
\newcommand{\mc}[1]{\mathcal{#1}}

\def\M{{\bar{M}}}
\def\N{{\bar{N}}}
\def\ph{{\bar{\phi}}}
\def\W{{\hat{W}}}
\def\K{{\hat{K}}}
\def\la{{\hat{\lambda}_6}}
\def\Z{\hat{Z}}
\def\n{\bar{n}}
\def\m{\bar{m}}
\def\l{\lambda}
\def\pH{\partial_{h}}
\def\pHb{\partial_{\bar{h}}}

\def\ä{\"{a}}

\def\p{\phi}
\def\po{\phi_0}

\title{Supergravity origin of the MSSM inflation}

\author{
Kari Enqvist$^{1,2,}$\footnote{E-mail: kari.enqvist@helsinki.fi}~,
Lotta Mether$^{1,2,}$\footnote{E-mail: lotta.mether@helsinki.fi}~
and Sami Nurmi$^{2,}$\footnote{E-mail: sami.nurmi@helsinki.fi}~
\\
${}^1$ Helsinki Institute of Physics, P.O. Box 64, FIN-00014 University of Helsinki, Finland\\
${}^2$ Department of Physical Sciences, P.O. Box 64, FIN-00014
University of Helsinki, Finland}

\abstract{We consider the supergravity origin of the recently
proposed MSSM inflationary model, which relies on the existence of a
saddle point along a dimension six flat direction. We derive the
conditions that the K\"{a}hler potential has to satisfy for the
saddle point to exist irrespective of the hidden sector vevs. We
show that these conditions are satisfied by a simple class of
K\ähler potentials, which we find to have a similar form as in
various string theory compactifications. For these potentials, slow
roll MSSM inflation requires no fine tuning of the soft
supersymmetry breaking parameters.}

\preprint{HIP-2007-33/TH}


\maketitle

\begin{document}

%%%%%%%%%%%%%%%%%%%%%%%%%%%%%%%%%%%%%%%%%%%%%%%%%%%%%%%%%%%%%%%%%%%%%%%%%%%%%%
%                         NEW SECTION
%%%%%%%%%%%%%%%%%%%%%%%%%%%%%%%%%%%%%%%%%%%%%%%%%%%%%%%%%%%%%%%%%%%%%%%%%%%%%%

\section{Introduction}
Although conventionally inflation is assumed to take place at a very
high energy scale and be driven by the slow roll motion of an
unknown singlet field, the inflaton, recently it has been pointed
out \cite{AEGM, AEGJM} that inflation can in fact be realized
already within the Minimally Supersymmetric Standard Model (MSSM).
In this case the inflaton is a gauge invariant combination of squark
or slepton fields. The flatness of the inflaton potential is
provided by supersymmetry and the gauge symmetries, which together
give rise to about 300 flat directions in the space of scalar fields
\cite{DRT} (for a review of the physics of the MSSM flat directions,
see~\cite{KARI-REV}). Along these flat directions, in the limit of
exact supersymmetry (susy), the scalar potential vanishes
identically. However, the flat directions are lifted by
non-renormalizable superpotential terms, as well as by soft susy
breaking \cite{GKM}; of particular importance for MSSM inflation are
the non-renormalizable $A$-terms $\propto W_n$, where $W_n$ is the
non-renormalizable superpotential of dimension $n$. The
dimensionality of the non-renormalizable terms depends on the
particular direction, as does the existence of the $A$-terms, which
are absent for some directions.

As discussed in \cite{AEGM}, phenomenologically acceptable slow roll
MSSM inflation can arise along the dimension six flat directions
{\bf udd} and {\bf LLe}, which we denote by the field $\Phi$. The
flat direction field is complex, and in the complex plane there
exists a set of discrete directions along which the contribution of
the $A$-term is most negative. Along these directions the MSSM
inflaton potential reads
\beq \label{scpot} V = \frac{1}{2}m^2\p^2 - \frac{A\l}{6}\p^6 +
{\l}^2 \p^{10}\ ,
\eeq
where $\phi$ is the absolute value of the
field, $m$ and $A$ are the soft susy breaking terms, $\lambda$ is an
effective coupling constant and we have set $M_P\equiv1$.
Generically, the potential Eq.~(\ref{scpot}) does not as such give
rise to inflation. However, one may notice that it has a secondary
minimum at %%
\beq
\label{po}
\po=\Big(\frac{A}{20\l}\Big)^{1/4}\ll1\ ,
\eeq
which becomes a saddle point if the condition
\beq\label{saddlecond}
A^2=40m^2
\eeq
holds. In that case the potential is extremely flat with
$V'(\phi_0)=V''(\phi_0)=0$. In the vicinity of the saddle point the
potential is given by %%
\beq
\label{saddlepot}
V(\p)\approx
V(\po)+\frac{1}{6}V'''(\po)(\p-\po)^3=V(\po)+\frac{16}{3}\frac{m^2}{\po}(\p-\po)^3\
.
\eeq
If in the initial state $\phi\simeq \phi_0$, there follows a period
of slow roll inflation with a very low scale of $H_{inf}\sim
1-10$~GeV, assuming $\l\sim{O}(1)$, and a spectral index of $n\simeq
0.92$ \cite{AEGM}. Slight deviations from the saddle point condition
Eq.~(\ref{saddlecond}) modify the spectral index somewhat (see
\cite{AEGJM}). Because of the low inflationary scale, there are no
observable tensor perturbations.

The great virtue of the MSSM inflation is that the inflaton
couplings to Standard Model particles are known and, at least in
principle, measurable in laboratory experiments such as LHC or a
future Linear Collider. The inflaton mass is directly related to the
slepton or squark masses and the model can thus be tested in the
laboratory.

However, the obvious disadvantage is the fine tuning implicit in the
saddle point condition Eq.\ (\ref{saddlecond}). Slow roll
inflation\footnote{For a discussion of the inflationary properties
of a potential that has the generic form of Eq.~(\ref{scpot}), see
\cite{LythDimopoulos}; for a discussion on dark matter and MSSM
inflation, see \cite{Allahverdi:2007vy}.} requires that the ratio
$A/m$ should be tuned to the saddle point with an accuracy of about
$10^{-16}$; otherwise the slow roll properties of the potential
Eq.~(\ref{scpot}) would be spoiled \cite{AEGJM}. Since in the MSSM
the soft susy breaking parameters are put in by hand, there can be
no explanation for the saddle point condition other than simple
finetuning. Thus the relation Eq.~(\ref{saddlecond}) must reflect
physics that is beyond the MSSM and in particular the mechanism of
supersymmetry breaking. Hence the values of the soft susy breaking
parameters reflect the properties of the hidden sector. The question
then is: is it possible to realize the saddle point condition
naturally in some supergravity model as defined by the K\"ahler
potential? This means that the condition Eq.~(\ref{saddlecond})
should not be just an accidental coincidence that emerges when the
hidden sector fields settle in their vevs, but rather a generic
condition that holds irrespective of the hidden sector field values.
In the present paper we demonstrate that this is indeed the case.
Moreover, the form of the K\"ahler potential turns out to be rather
suggestive, with features that can be found in certain string
theoretical compactification schemes.

%%%%%%%%%%%%%%%%%%%%%%%%%%%%%%%%%%%%%%%%%%%%%%%%%%%%%%%%%%%%%%%%%%%%%%%%%%%%%%

\section{The scalar potential}

Our aim is to identify a class of K\"ahler potentials that generate
such soft susy breaking terms for the flat direction $\Phi$
that the saddle point condition (\ref{saddlecond}) is identically
satisfied. It is obvious that the simplest, flat K\"ahler potential
will not do the job; instead, one has to consider more complicated possibilities.
We focus on soft terms generated through F-term
susy breaking (recall that the flat directions are D-flat also
in supergravity). In this case the scalar potential is
determined solely by the function
 \beq
  G(\Phi_M, \Phi^*_M) = K(\Phi_M, \Phi^*_M) + \log|W(\Phi_M)|^2\ ,
 \eeq
where $K$ and $W$ are respectively the K\"ahler potential and the
superpotential. Here $\Phi_M$, which includes both the hidden sector
fields $h_m$ and the flat direction inflaton field $\Phi$, denotes
the scalar part of the corresponding chiral superfield. Assuming
vanishing D-terms also in the hidden sector, the tree-level scalar
potential reads
 \baq \label{V(G)}
  V & = & e^G \Big( G^{M\N} G_M G_\N - 3 \Big)\ ,
 \eaq
where the lower indices $_M$ and $_\M$ refer to derivatives with
respect to $\Phi_M$ and $\Phi^*_M$, and the matrix $G^{M\N}=K^{M\N}$
is the inverse of the K\"ahler metric $G_{M\N}=K_{M\N}$.

For the dimension $6$ flat directions that we are considering as the
inflaton, the superpotential is of the form
 \beq
 \label{W}
  W = \W + \frac{\la}{6}\Phi^6 \equiv \W + I\ ,
 \eeq
where $I$ is the lowest order non-renormalizable term that lifts the
flat direction. The superpotential may also contain all possible
higher order terms allowed by symmetries but these will not affect
our analysis and have therefore been suppressed. In Eq.~(\ref{W})
and elsewhere in the text, we use the hat to denote quantities that
are independent of $\Phi$, but are functions of the hidden sector
fields. This is in general the case for the $\Phi$-independent term
$\W$ of the superpotential, as well as for the coupling constant
$\la$ of the non-renormalizable term. However, since our focus is on
finding a K\"ahler potential that satisfies the relation
Eq.~(\ref{saddlecond}), we will neglect the hidden sector dependence
of the superpotential, and hence treat these quantities as constants
throughout this paper. In this context, it is worth noting that, in
order to ensure the validity of the MSSM inflation scenario, we are
assuming the flat direction to be the only dynamical variable during
inflation. Thus we are implicitly requiring that the hidden sector
fields are stabilized before the beginning of inflation either by
the neglected superpotential terms or through some other
mechanism.\footnote{The inclusion of the hidden sector fields, even
when stabilized before the onset of inflation, may in general lead
to additional fine-tuning conditions on the inflationary potential.
A detailed analysis of these effects however requires precise
knowledge of the nature and dynamics of the hidden sector fields and
is beyond the scope of this paper. A qualitative discussion of this
issue can be found in Ref.~\cite{Ellis:2006ara}, even if the results
therein as such are not applicable here.}

Given the superpotential Eq.~(\ref{W}), the scalar potential
Eq.~(\ref{V(G)}) can be written as
 \beq
 \label{V}
  V = |\W|^2f + \W I^*g + \W^*Ig^* + |I_\phi|^2k\ ,
 \eeq
where
 \baq
 \label{f}
  f & = & e^K \Big(K^{M\N}K_MK_\N - 3\Big)\ ,  \\
 \label{g}
  g & = & e^K \Big(\frac{6}{\Phi^*}K^{M\ph}K_M + K^{M\N}K_MK_\N - 3\Big)\ ,\\
 \label{k}
  k & = & e^K \Big(K^{\phi\ph} + \frac{\Phi}{6}K^{M\ph}K_M + \frac{\Phi^*}{6}K^{\phi\M}K_\M +
   \frac{\Phi\Phi^*}{36}(K^{M\N}K_MK_\N-3)\Big)\ .
 \eaq

To find the explicit expression for the potential Eq. $(\ref{V})$,
one needs to determine the K\ähler potential. Here we consider
K\"{a}hler potentials of the generic perturbative form
\beq
\label{K}
K=\K+\Z_2\p^2+\Z_4\p^4+\Z_6\p^6+\ldots\ ,
\eeq
where $\p$ denotes the absolute value, $\Phi=\phi~{\rm
exp}(i\theta)$. Using the K\ähler potential Eq.~(\ref{K}) to expand
the coefficients $f,g$ and $k$ in Eq.~(\ref{V}) in powers of $\p$
and keeping only the lowest order terms, the scalar potential
Eq.~(\ref{V}) becomes %%
\beq
\label{leading}
V=V_0+V_2\p^2+V_6\p^6+V_{10}\p^{10}\ ,
\eeq
where
\baq
\label{V0}
V_{0~}&=&e^{\K}|\W|^2\Big(\K^{m}\K_{m}-3\Big)\ ,\\
\label{V2}
V_{2~}&=&e^{\K}|\W|^2\Z_2\Big(\K^m\K_m+\K^m\K^{\bar{n}}(\Z_2^{-2}\Z_{2m}\Z_{2\bar{n}}
-\Z_2^{-1}\Z_{2m\bar{n}})-2\Big)\ ,\\
\label{V6}
V_{6~}&=&e^{\K}|\W||\la|{\rm
cos}(\xi-6\theta)\Big|\frac{1}{3}\K^m\K_m-2\Z^{-1}_2\K^{\m}\Z_{2\m}+1\Big|\ ,\\
\label{V10}
V_{10}&=&e^{\K}|\la|^2\Z^{-1}_2\ ,
\eaq
the phase $\xi$ in $V_6$ reads
\baq\label{theta} \xi&\equiv&{\rm
arg}\Big(\frac{1}{6}\K^m\K_m-\Z^{-1}_2\K^{\m}\Z_{2\m}+\frac{1}{2}\Big)+{\rm
arg}(\W)-{\rm arg}(\la)
\eaq
and indices are raised and lowered with $\K^{M\N}$ and $\K_{M\N}$
respectively. Here $V_0$, $V_6$ and $V_{10}$ result from the leading
order expansion of $f,g$ and $k$ respectively, whereas $V_2$ is
obtained by expanding $f$ to next to leading order. The expansion is
performed in this manner since the constant $V_0$, which would give
rise to a cosmological constant, will be neglected
henceforth\footnote{We assume the cosmological constant to be
adjusted to the observationally required value either by terms
arising from the hidden sector dependent superpotential, or by some
other (yet unknown) mechanism.}. Thus $V_2$ becomes the leading
non-trivial term in the expansion of $f$ and Eq.~(\ref{leading})
with $V_0$ removed then constitutes the leading order potential. In
the following, the term leading order will be understood precisely
in this sense, i.e. as leading non-trivial order.

By expanding $f,g$ and $k$ in Eq.~(\ref{V}) to next to leading
order, one finds a first order correction $\Delta_1V$ to the
potential Eq.~(\ref{leading}), at next to next to leading order one
finds a second order correction $\Delta_2V$, and so on. In carrying
out this sort of expansion, we implicitly restrict our analysis to
the values of $\p$ for which all the terms in Eq.~(\ref{V}) are
comparable, which will certainly be the case in the vicinity of the
eventual saddle point. Using Eqs. (\ref{V}) -- (\ref{K}), an order
of magnitude approximation of the $n$-th order correction to the
leading order potential is then given by %%
\beq
\label{Vn}
\Delta_n V\sim e^{\K}|\W|^2\Z^{n+1}_2\p^{2n+2}\ .
\eeq
%%

%%%%%%%%%%%%%%%%%%%%%%%%%%%%%%%%%%%%%%%%%%%%%%%%%%%%%%%%%%%%%%%%%%%%%%%%%%%%%%%%%%%%%%%%%%%%%%%%%%%%%%%%

\section{The saddle point condition}

%%%%%%%%%%%%%%%%%%%%%%%%%%%%%%%%%%%%%%%%%%%%%%%%%%%%%%%%%%%%%%%%%%%%%%%%%%%%%%%%%%%%%%%%%%%%%%%%%%%%%%%%

In this Section we consider the restrictions placed by the saddle
point condition Eq.~(\ref{saddlecond}) on the leading order
potential. The role of higher order corrections will be discussed in
the next Section.

By choosing the phase $\theta$ of $\Phi$ such that ${\rm
cos}(\xi-6\theta)=-1$ in Eq.~(\ref{V6}), the $\theta$ dependent part
of the leading order potential Eq.~(\ref{leading}) is minimized and
we recover Eq.~(\ref{scpot}), where %%
\baq
\label{m}
m^2&=&2e^{\K}|\W|^2\Z_2\Big(\K^m\K_m+\K^m\K^{\bar{n}}(\Z_2^{-2}\Z_{2m}\Z_{2\bar{n}}
-\Z_2^{-1}\Z_{2m\bar{n}})-2\Big)\ ,\\
\label{A}
A&=&e^{\K/2}|\W|\Z_2^{1/2}\Big|2\K^m\K_m-12\Z^{-1}_2\K^{\m}\Z_{2\m}+6\Big|\ ,\\
\label{c}
\l^2&=&e^{\K}|\la|^2\Z^{-1}_2\ .
\eaq
The saddle point condition Eq.~(\ref{saddlecond}) then becomes
\baq
\label{saddle}
|\K^m\K_m-6\Z_2^{-1}\K^{\m}\Z_{2\m}+3|^2&=&20(\K^m\K_m+
\K^m\K^{\bar{n}}(\Z_2^{-2}\Z_{2m}\Z_{2\bar{n}}-\nonumber\\&&
\Z_2^{-1}\Z_{2m\bar{n}})-2)\ ,
\eaq
which is a partial differential equation for two unknown functions,
$\K$ and $\Z_2$.

As a simple example we first consider a scenario in which there is
only one hidden sector field $h$. Treating the functions $\K$ and
$\Z_2$ as independent variables, Eq.~(\ref{saddle}) implies
\beq
\pH\K\pHb\K=-\beta\pH\pHb\K\ ,
\eeq
where $\beta$ is a constant. An analogous equation appears in
no-scale supergravity models \cite{noscale} and is solved for
\beq
\label{1khat}
\K=\beta{\rm log}(h+h^{*})\ .
\eeq
Using this result and assuming $\Z_2=\Z_2(h+h^{*})$\ , the saddle
point condition Eq.~(\ref{saddle}) in the one-dimensional case
becomes %%
\beq
(3-\beta+6(h+h^{*})\pH {\rm
log}Z_2)^2=20(-\beta-2-(h+h^{*})^2\pH^2{\rm log}\Z_2)\ ,
\eeq
whose general solution is
\beq
\Z_2=(h+h^{*})^{-2/9+\beta/6}\Big[
c_1(h+h^{*})^{\omega(\beta)}+c_2(h+h^{*})^{-\omega(\beta)}\Big]^{5/9}\
,
\eeq
where $c_1,c_2$ are constants, and
$\omega(\beta)=1/2\sqrt{-17-6\beta}$ . The solution takes a
particularly simple form if one of the constants $c_1,c_2$ is zero.

To find a solution of Eq.~(\ref{saddle}) in the general case with
several hidden sector fields, we make an Ansatz motivated by the
one-dimensional case and write
\beq
\label{k2}
K=\sum_m\beta_m{\rm
log}(h_m+h_{m}^{*})+\kappa\prod_m(h_m+h_{m}^{*})^{\alpha_m}\p^2+\mc{O}(\p^4)\
,
\eeq
where $\alpha_m,\beta_m$ and $\kappa$ are constants. K\"{a}hler
potentials of this type are found e.g. in Abelian orbifold
compactifications of heterotic string theory \cite{abelorbi} as well
as in intersecting D-brane models \cite{stringmodels}. In both cases
the moduli fields play the role of the hidden sector fields $h_m$.
Here we will, however, treat the parameters in Eq.~(\ref{k2}) from a
phenomenological point of view, without any particular string
scenario in mind.

The Ansatz Eq.~(\ref{k2}) solves Eq.~(\ref{saddle}) provided the
parameters are related by
\beq
\label{ab}
\alpha(36\alpha+16-12\beta)+(\beta+7)^2=0\ ,
\eeq
where
\baq
\alpha&=&\sum_m\alpha_m\ ,\\
\beta&=&\sum_m\beta_m\ .
\eaq
In Table~1 we list solutions to Eq.~(\ref{ab}) for which the soft
susy breaking terms are nonzero and the $\alpha_m$ are rational
numbers. In the string context $-\beta$ generically measures the
number of hidden sector fields and therefore we restrict ourselves
to the lowest values of $\beta$.
\begin{table}[h!]
\begin{center}
\label{table1} \caption{Values of $\beta$ and $\alpha$ in the
K\"ahler potential Eq.~$(3.9)$ for which the saddle point condition
is satisfied identically.} \vspace{12pt}
\begin{tabular}{|c||c|}
\hline
$\beta=\sum_m\beta_m$ & $\alpha=\sum_m\alpha_m$ \\
\hline
$-3 $ & $-\frac{4}{9}$\\
\hline
$-7$ & $~~0$  \\
\hline $-7$ & $-\frac{25}{9}$ \\
\hline $-11$ & $-\frac{1}{9}$ \\
\hline $-11$ & $-4$ \\
\hline
\end{tabular}
\end{center}
\end{table}

To summarize, for the leading order potential the saddle point
condition Eq.~(\ref{saddlecond}) is satisfied identically with the
class of K\ähler potentials determined by Eq.~(\ref{k2}) and the
conditions on the parameters $\beta_m,\alpha_m$ as given in Table~1.
While there definitely exist other solutions of Eq.~(\ref{saddle})
as well, Eq.~(\ref{k2}) represents the only class of solutions for
which $\Z_2$ is separable and the hidden sector dependence is of
similar functional form for all the fields, provided that
$K=K(h_m+h_m^{*})$ and the hidden sector metric $\K_{m\n}$ is
diagonal. Since we are not making any specific assumptions about the
physical nature of the hidden sector fields, these seem to be quite
natural conditions to impose on $\Z_2$.

%%%%%%%%%%%%%%%%%%%%%%%%%%%%%%%%%%%%%%%%%%%%%%%%%%%%%%%%%%%%%%%%%%%%%%%%%%%%%%%%%%%%%%%%%%%%%%%%%%%%%%%%

\section{Higher order corrections}

%%%%%%%%%%%%%%%%%%%%%%%%%%%%%%%%%%%%%%%%%%%%%%%%%%%%%%%%%%%%%%%%%%%%%%%%%%%%%%%%%%%%%%%%%%%%%%%%%%%%%%%%

In the vicinity of the saddle point as given by Eq.~(\ref{po}), the
slope of the leading order potential Eq.~(\ref{scpot}) is extremely
small. Therefore, one may ask whether the corrections arising from
the expansion of the potential Eq.~(\ref{V}) to higher orders will
destroy this flatness. In this Section, we show that the required
flatness \cite{AEGM, AEGJM} is maintained if, in addition to the
leading order potential, also the first and second order corrections
satisfy certain conditions. Analogously to the leading order
results, we find a form of the K\"{a}hler potential for which all
these conditions are satisfied identically, i.e. irrespective of the
vevs of hidden sector fields.

Within the slow roll approximation, the dynamics are determined by
the first derivative of the potential. Therefore the condition for
the $n$-th order correction, $\Delta_n V$, not to alter the leading
order results can be expressed as %%
\beq
\label{delta_condition}
\Delta_n V'(\p)\ll V'(\p)\sim
10^{-3}\frac{m^2\hat{Z}_2^2}{N(\p)^2}\po^5\ ,
\eeq
where the derivative of the leading order potential, $V'(\p)$, has
been written in terms of the e-foldings $N(\p)$ remaining until the
end of inflation \cite{AEGM, AEGJM}. Using Eq.~(\ref{Vn}), this
condition becomes
\beq
\label{n-th_condition}
\Z_2^{n-2}\po^{2n-4}\ll 10^{-3}N(\p)^{-2}\ ,
\eeq
which is satisfied automatically for $n>2$ since\footnote{Note that
it is actually the canonically normalized field $\p_{\rm
can}\sim\Z_2^{1/2}\p$ that is the MSSM inflaton. Therefore
$\Z_2^{1/2}\p\ll1$. However, $V(\p_{\rm can})$ has a saddle point
under the same conditions as $V(\p)$ and the choice of the field
variable plays no role in our analysis.} $\Z_2^{1/2}\po\ll1$. This
means that the third and higher order corrections are negligible and
require no further attention. The corrections $\Delta_1V$ and
$\Delta_2V$, on the other hand, can not be made small simply by
adjusting parameters, but one needs to set their derivatives to zero
identically. To be more precise, Eq.~(\ref{n-th_condition}) is
satisfied if $\Delta V_1'(\po)=\Delta V_1''(\po)=0$ and $\Delta_2
V'(\po)=0$.

The first order corrections to the leading order potential can be
written as
\beq\label{dV1} \Delta_1 V=V_4\p^4+V_8\p^8+V_{12}\p^{12}\ ,
\eeq %%
where the coefficients are obtained from Eqs. (\ref{V}) -- (\ref{k})
by retaining only the next to leading order terms. With the leading
order K\"{a}hler potential given by Eq.~(\ref{k2}), the conditions
$\Delta V_1'(\po)=\Delta V_1''(\po)=0$ yield a pair of partial
differential equations for the coefficient $\hat{Z}_4$ in the
K\"{a}hler potential Eq.~(\ref{K}) whose only solution is
\beq\label{z4}
\hat{Z}_4=\mu(\alpha,\beta,\gamma(\alpha,\beta))\hat{Z}_2^2\ ,
\eeq
where $\gamma=\sum_m\alpha_m^2/\beta_m$ and we have assumed
$\Z_4=\Z_4(h_m+h_{m}^{*})$. The parameters $\mu$ and $\gamma$ are
not freely selectable but completely determined by $\alpha$ and
$\beta$ such that $\Delta V_1'(\po)=\Delta V_1''(\po)=0$. In Table~2
below we give their values for the choices of $\alpha$ and $\beta$
considered in this work.

In a similar manner, the second order correction reads
\beq
\label{dV2}
\Delta_2 V=V_6\p^6+V_{10}\p^{10}+V_{14}\p^{14}\ ,
\eeq
with $V_6$, $V_{10}$ and $V_{14}$ here denoting the next to next to
leading order part of Eqs. (\ref{V}) -- (\ref{k}). In this case, the
conditions to be placed on the K\"{a}hler potential are less
stringent since one only needs to set $\Delta V_2'(\po)=0$. Assuming
$\hat{K},\hat{Z}_2,\hat{Z}_4$ in Eq.~(\ref{K}) to be given by Eqs.
(\ref{k2}), (\ref{z4}), the condition $\Delta V_2'(\po)=0$ is
satisfied for
\beq
\label{4th_condition}
\hat{Z}_6=\nu(\alpha,\beta,\delta)\hat{Z}_2^3\ ,
\eeq
where $\delta=\sum_m\alpha_m^3/\beta_m^2$, and only the relation
between $\delta$ and $\nu$ is determined by $\alpha$ and $\beta$,
see Table~2 below. Moreover, in addition to $\hat{Z}_6$ given by
Eq.~(\ref{4th_condition}), the $\mc{O}(\p^6)$ part of the K\"{a}hler
potential may also contain solutions of the homogeneous part of the
partial differential equation arising from $\Delta_2 V'(\po)=0$.

Thus we have shown that there exists a class of K\ähler potentials
for which the extreme flatness of the MSSM inflaton potential
\cite{AEGM, AEGJM} is generated and maintained also in the presence
of higher order corrections irrespective of the hidden sector vevs.
This class of K\ähler potentials can be written in the form
\baq
\label{kahler}
K&=&\sum_m\beta_m{\rm
log}(h_m+h_m^{*})+\kappa\prod_m(h_m+h_m^{*})^{\alpha_m}\p^2+\mu\Big(\kappa\prod_m(h_m+h_m^{*})^{\alpha_m}\Big)^2\p^4
+\nonumber\\&&
\nu\Big(\kappa\prod_m(h_m+h_m^{*})^{\alpha_m}\Big)^3\p^6+\mc{O}(\p^8)\
,
\eaq
where the parameters are subject to the conditions given in Table 2
below.
\begin{table}[h!]
\begin{center}
\label{table}
\caption{The coefficients of the higher order terms in the
K\"{a}hler potential that guarantee the flatness of the MSSM
inflaton potential.}
\vspace{12pt}
\begin{tabular}{|c||c||c||c||c||c|}
\hline
$\rule{0pt}{3ex}\beta=\sum_m\beta_m$ & $\alpha=\sum_m\alpha_m$ & $\gamma=\sum_m\alpha_m^2/\beta_m$ & $\mu$ & $\delta=\sum_m\alpha_m^3/\beta_m^2$ & $\nu$\\
\hline
$\rule{0pt}{2.5ex}-3 $ & $-\frac{4}{9}$ & $~~\frac{1}{9}$ & $-\frac{7}{36}$ & $\frac{91}{324}$& $\nu$\\
\hline
$\rule{0pt}{2.5ex}-7$ & $~~0$ & $~~0$ & $~~\frac{1}{12}$ & $\delta$ & $\nu$\\
\hline
$\rule{0pt}{2.5ex}-7$ & $-\frac{25}{9}$ & $-\frac{10}{9}$ &
$~~\frac{2}{9}$ & $-\frac{1654}{1863}+\frac{162}{23}\nu$ & $\nu$\\
\hline
$\rule{0pt}{2.5ex}-11$ & $-\frac{1}{9}$ & $~~\frac{1}{21}$ &
$~~\frac{13}{126}$ & $-\frac{8465}{75411}+\frac{162}{19}\nu$ & $\nu$ \\
\hline
$\rule{0pt}{2.5ex}-11$ & $-4$ & $-\frac{29}{21}$ &
$~~\frac{17}{84}$ & $-\frac{2491}{2940}+\frac{36}{5}\nu$ & $\nu$ \\
\hline
\end{tabular}
\end{center}
\end{table}

%%%%%%%%%%%%%%%%%%%%%%%%%%%%%%%%%%%%%%%%%%%%%%%%%%%%%%%%%%%%%%%%%%%%%%%%%%%%%%%%%%%%%%%%%%%%%%%%%%%%%%%%
\newpage

\section{Conclusions}

In this work we have considered the supergravity origin of the
recently proposed MSSM inflationary model \cite{AEGM, AEGJM}. In
particular, we have shown that for the simple class of K\ähler
potentials given by Eq. (\ref{kahler}), the extremely flat inflaton
potential is produced identically in F-term supersymmetry breaking.
The desired form of the potential is thus obtained for all hidden
sector vevs and not just for some carefully chosen vacua.

The class of K\ähler potentials Eq. (\ref{kahler}) has a number of
appealing features. Firstly, although it is necessary to fix the
potential up to $\mc{O}(\p^6)$, no new functions need to be
introduced in addition to $\K$ and $\Z_2$ appearing already in the
leading order expression, Eq. (\ref{k2}). Moreover, it is
interesting to note that the form of K\"ahler potentials for which
the MSSM inflationary scenario happens to be realized, is very
common in string theory compactifications. As mentioned in Section
3, K\"ahler potentials of the form given in Eq. (\ref{k2}) arise
e.g. in Abelian orbifold compactifications of the heterotic string
theory \cite{abelorbi} and in intersecting D-brane models
\cite{stringmodels}. In the heterotic case, the parameters
$\alpha_m$ are modular weights, whereas in the intersecting brane
models they depend on internal fluxes of the branes. To our
knowledge, there is no specific compactification known so far, which
would generate exactly the required values given in Table~1. This
certainly would be a matter worth further investigation, especially
keeping in mind that none of the string theory compactifications
known to date produce exactly the actual MSSM. It would be very
interesting to find a compactification that generates the saddle
point along a d=6 flat direction of the MSSM.

The argumentation may also be turned the other way around. A
supergravity model with F-term supersymmetry breaking, an MSSM like
visible sector and a K\ähler potential of the form given in Eq.
(\ref{kahler}) may naturally lead to an inflationary period driven
by the MSSM degrees of freedom and with properties consistent with
the observed cosmological data \cite{data}, provided the initial
condition is such that the flat direction field finds itself in the
vicinity of the saddle point. At this point, however, we wish to
emphasize that since we are assuming the flat direction to be the
only dynamical degree of freedom during inflation, we are also
implicitly assuming the moduli fields to be stabilized before the
beginning of inflation. Although the exceptionally low scale of
inflation gives some justification for this assumption, the validity of it
is highly model dependent and non-trivial, and should be discussed
separately in the context of any realistic supergravity model.
In any case, while inflation may still be possible even if the moduli
fields are not stabilized, the resulting inflationary model will 
in general be different from the MSSM inflation discussed in this paper.

Finally, the supergravity model leading to the MSSM inflation can,
at least in principle, be tested not only by cosmological
observations but also in particle accelerators. For instance, given
the K\ähler potential Eq.~(\ref{kahler}), for a non-flat direction
$\psi$ with a renormalizable superpotential $W(\psi)=\frac 13
\hat\lambda_3 \psi^3$ one finds that the trilinear A-term is given
by $A_3=m_{\psi}\sqrt{2}(\alpha-\beta/3){\rm
cos}\,\xi/\sqrt{\alpha-\beta -2}$~, where ${\rm cos}\,\xi$ contains
the phase information. Once scaled down to LHC energies by the
renormalization group equations, such relations have obvious
ramifications for both sparticle phenomenology and inflation.

%%%%%%%%%%%%%%%%%%%%%%%%%%%%%%%%%%%%%%%%%%%%%%%%%%%%%%%%%%%%%%%%%%%%%%%%%%%%%%%%%%%%%%%%%%%%%%%%%%%%%%%%
%acknowledgements
%%%%%%%%%%%%%%%%%%%%%%%%%%%%%%%%%%%%%%%%%%%%%%%%%%%%%%%%%%%%%%%%%%%%%%%%%%%%%%%%%%%%%%%%%%%%%%%%%%%%%%%%

\acknowledgments{KE is partially supported by the Academy of Finland
grant 114419. LM is supported by the Vilho, Yrj\"o and Kalle
V\"ais\"al\"a Foundation and SN by the Graduate School in Particle
and Nuclear Physics. This work was also partially supported by the
EU 6th Framework Marie Curie Research and Training network
``UniverseNet'' (MRTN-CT-2006-035863).}

\end{document}